# An encoded asymmetric ligand for metal-selective topological assembly of two-dimensional metal–organic frameworks

Huimin Qi[1†], Jinkun Guo[1†], Xinyan Wu[1,2†], Weishan Li[1], Tongyang Zhao[1], Ze-Fan Yao[3], Hao Chen[1,4], Ling Zhang[1], Bin Jiang[5], Yi Liu[1], Haoyang Zhang[1], Yunlong Fan[1], Tianyang Chen[6], Qingqing Ji[2*], and Jin-Hu Dou[1*]

[1]National Key Laboratory of Advanced Micro and Nano Manufacture Technology; Key Laboratory of Polymer Chemistry and Physics of Ministry of Education, School of Materials Science and Engineering, Peking University, Beijing, 100871, China.

[2] School of Physical Science and Technology, ShanghaiTech University, Shanghai, 201210, China.

[3]Beijing National Laboratory for Molecular Sciences (BNLMS); Key Laboratory of Polymer Chemistry and Physics of Ministry of Education; Center of Soft Matter Science and Engineering, College of Chemistry and Molecular Engineering, Peking University, Beijing 100871, China.

[4] Key Laboratory of Artificial Structure and Quantum Control, Ministry of Education, School of Physics and Astronomy, Shanghai Jiao Tong University, Shanghai, 200240, China.

[5]National Key Laboratory of Advanced Micro and Nano Manufacture Technology; Beijing Advanced Innovation Center for Integrated Circuits, School of Integrated Circuits, Peking University, Beijing, 100871, China.

[6]School of Science and Engineering, The Chinese University of Hong Kong (Shenzhen), Longgang, Shenzhen, Guangdong, 518172, China.

†These authors contribute equally.

* Corresponding author. Email: doujinhu@pku.edu.cn (J. D.); Email: jiqq@shanghaitech.edu.cn (Q. J.).

*KEYWORDS: Encoded asymmetric ligand; 2D MOFs; Metal-selective topological assembly; Electronic properties; Reticular chemistry;*

**ABSTRACT:** Two-dimensional metal–organic frameworks (2D MOFs), with diverse topological architectures, provide a powerful platform for exploring unconventional electronic and lattice-dynamical responses. Yet their structural diversity remains fundamentally constrained by the fixed geometry of high-symmetry ligands. Here, we introduce an encoded asymmetric ligand, benzo[*b*]triphenylene-2,3,6,7,11,12-hexaol (BTH), for metal-selective topological assembly. By integrating multi-site coordination fields with sterically differentiated environments, BTH exhibits distinct topological programmability: different divalent metal ions direct divergent framework architectures. Specifically, coordination of BTH with divalent Cu(II) and Zn(II) ions assembles Cu-BTH-MOF with a dual-mode hexagonal pore topology and Zn-BTH-MOF with uniform hexagonal channels, respectively, as supported by PXRD Pawley refinement, structural simulations, and pore-size distribution analysis. Furthermore, this topological divergence is accompanied by a significant divergence in charge-transport properties, with Cu-BTH-MOF reaching an electrical conductivity of $1.186 \times 10^{-3}$ S $cm^{-1}$, more than six orders of magnitude higher than that of Zn-BTH-MOF ($3.38 \times 10^{-10}$ S $cm^{-1}$). This work establishes ligand desymmetrization as a programmable strategy for metal-selective topological diversification in 2D MOFs.

## Introduction

Manipulating the topological structures of two-dimensional (2D) materials can give rise to unconventional electronic and lattice phenomena—such as anisotropic charge transport, correlated electronic states, and collective vibrational behaviors.[1-3] Achieving such precise control in conventional crystalline materials is highly challenging, as rigid bonding constraints limit the available structural degrees of freedom. By comparison, 2D metal-organic frameworks (MOFs) offer a chemically programmable platform where coordination chemistry enables the systematic tuning of structures and functions, thereby providing a promising avenue for rational topological design.[4-6]

Network topology in conventional 2D MOFs is largely dictated by the geometry, symmetry, and connectivity of the organic ligand.[7-9] Ligands with high symmetry provide well-defined and repetitive coordination vectors, enabling reliable access to predictable architectures. However, this structural predictability also imposes an intrinsic limitation: the fixed ligand geometry commonly leads to very limited choice of framework topologies. A representative example is 2,3,6,7,10,11-hexahydroxytriphenylene (HHTP), a $D_{3h}$-symmetric ligand in which three equivalent catechol-type coordination directions define an isotropic 2-2-2 unit, where the numbers denote the number of fused phenylene rings extending from the central aromatic core along each coordination direction (**Figure 1a**-**b**). Propagation of this unit with divalent metals typically produces a uniform honeycomb lattice (**Figure 1c**).[10-12] Although synthetically robust, such deterministic assembly limits access to topologically diverse frameworks from a single molecular platform.

To expand the accessible topological space of MOFs, previous strategies have primarily relied on external chemical regulation, such as acid–base modulation, solvent effects, and coordinating additives, to bias framework assembly and induce alternative network formation.[13-16] In parallel, ligand desymmetrization has also been widely explored as an intrinsic design strategy to perturb equivalent coordination vectors and generate new topological outcomes. A variety of asymmetric linkers, including molecular hinge motifs,[17] low-symmetry scaffolds,[18-21] zigzag geometries,[22-23] and macrocyclic blocks,[24]have been developed to construct nontrivial frameworks. Representative examples include asymmetric linkers in the MOF-74 family that afford amorphous yet porous Mg-based frameworks,[25] and low-symmetry ligands such as THPQ ($C_2$), which assemble into 2D conductive Cu-MOFs with random in-plane tiling and high defect density.[26] Related concepts have also been extended to COFs, where desymmetrized vertex design enables periodically heterogeneous pore structures.[27] Collectively, these studies demonstrate that symmetry reduction is effective in generating new topologies. However, a modified ligand typically leads to a single alternative framework rather than enabling selective access to multiple topologies from a single system. Importantly, metal-dependent coordination selectivity has not been utilized as a design principle for directing framework topology.

Herein, we report an encoded asymmetric ligand, benzo[*b*]triphenylene-2,3,6,7,11,12-hexaol (BTH), for metal-selective topological assembly of 2D MOFs. BTH retains the hexadentate coordination motif of the parent triphenylene scaffold while introducing chemically non-equivalent hydroxyl groups and sterically differentiated coordination environments. This site differentiation enables different divalent metal ions to selectively access distinct assembly pathways from the same ligand precursor. Unlike the isotropic 2-2-2 motif of HHTP, which propagates exclusively through a single 2+2 connection mode, BTH exhibits an anisotropic 2-2-3 motif (**Figure 1d**), permitting multiple local connection modes, including 3+3, 2+3, *trans*-2+2, and *cis*-2+2 arrangements (**Figure 1e**). The added two numbers specify the phenylene arms from two BTH ligands being connected, reflecting the varying coordination vectors. Propagation of these motifs theoretically gives rise to diverse tiling patterns, such as [446], [455], mixed [444,46] and [555,45] networks (**Figure 1f**-**i**). Experimentally, coordination with Cu(II) produces Cu-BTH-MOF with a dual-mode hexagonal pore topology ([444,46]), whereas coordination with Zn(II) affords Zn-BTH-MOF with uniform hexagonal channels([446]). This one-ligand, divergent-topology behavior demonstrates that BTH functions as a programmable asymmetric building block, capable of encoding multiple local assembly pathways, providing a versatile platform for controlling 2D MOF topology. Furthermore, we show that these divergent topologies produce significant differences in charge transport properties, with a difference exceeding six orders of magnitude.

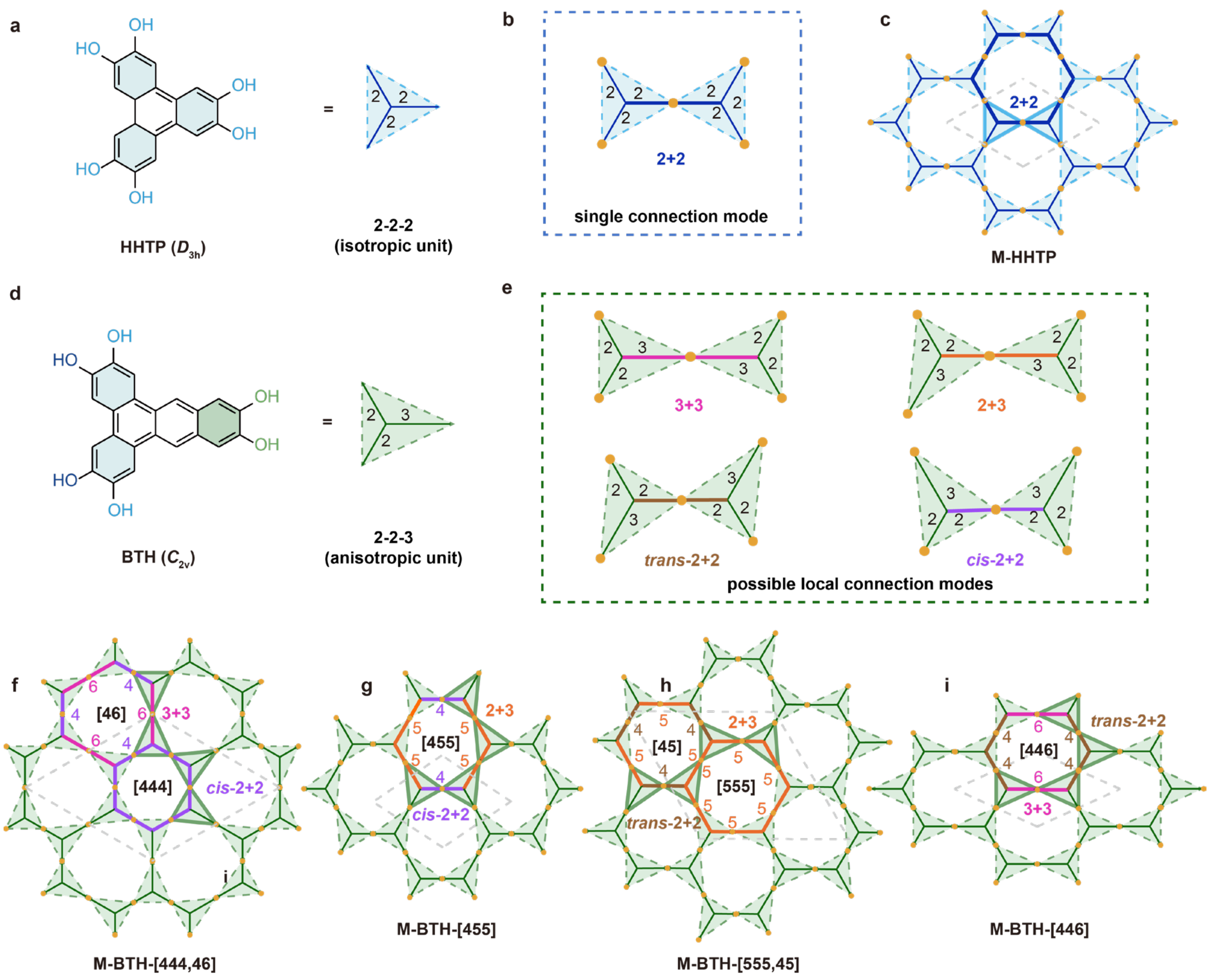

Figure 1. Schematic illustration of programmable asymmetric ligand design for metal-selective topological assembly in 2D MOFs. (a) HHTP ligand ($D_{3h}$) with six-equivalent OH groups; the isotropic HHTP-derived 2-2-2 coordination unit is shown as a blue triangle. (b) The isotropic 2-2-2 coordination unit leads to a single propagation mode, with metal nodes represented as yellow dots. (c) Propagation of the high-symmetric 2-2-2 unit with divalent metal ions leads to a uniform honeycomb lattice (M-HHTP). (d) BTH ligand ($C_{2v}$) with site-differentiated OH groups, giving the anisotropic BTH-derived 2-2-3 coordination unit shown as a green triangle. (e) The anisotropic 2-2-3 coordination unit of BTH enables multiple metal-directed local motifs, including 3+3 (pink), 2+3 (orange), *trans*-2+2 (brown), and *cis*-2+2 (purple) linkages. (f-i) Representative BTH-based tiling structures, including M-BTH-[444,46], M-BTH-[455], M-BTH-[555,45] and M-BTH-[446], derived from mixed 3+3, 2+3, *trans*-2+2, and *cis*-2+2 motifs. Enlarged green triangles highlight the coexistence of multiple local motifs within each framework.

## Results and Discussion

We designed a $C_{2v}$-symmetric ligand, benzo[*b*]triphenylene-2,3,6,7,11,12-hexaol (BTH), to construct topologically divers 2D MOFs. BTH was synthesized through Pd-catalyzed coupling,[28] $FeCl_3$-mediated oxidative cyclization,[29] and $BBr_3$-promoted demethylation,[30] with full characterization provided in the Supporting Information (**Schemes S1-S2** and **Figures S1-S8**). Compared to HHTP, BTH bears a single benzo-fused extension at one edge of the triphenylene core. This unique structural design retains all six hydroxyl coordination sites while lowering the molecular symmetry from $D_{3h}$ to $C_{2v}$. More importantly, it transforms the originally equivalent coordination environment of HHTP into site-differentiated hydroxyl domains with distinct electronic and steric characteristics. Thus, whereas HHTP behaves as a highly symmetric linker that typically directs divalent metals into a uniform honeycomb lattice, BTH provides an anisotropic, site-differentiated organic scaffold that serves as an encodable platform for metal-selective topological diversification in MOFs.

Consistent with this, density functional theory (DFT) calculations reveal a clear three-tier p$K_a$ distribution for BTH (7.93, 8.48, and 8.07), in stark contrast to the single p$K_a$ value of HHTP (8.71). The most acidic site of BTH (p$K_a$ 7.93, located on the benzo-fused edge) exhibits the lowest deprotonation energy barrier and therefore serves as the kinetically favored anchor point for triggering network growth (**Figure 2a**). We note that if all hydroxyl groups are fully deprotonated under the reaction conditions, the relative coordination strengths of different sites may also be affected by their intrinsic basicity and electronic structures; however, in the present system, we favor a partially deprotonated or kinetically controlled early-stage assembly scenario, in which the lower-p$K_a$ benzo-fused edge is more likely to participate preferentially in coordination. Furthermore, BTH displays a higher HOMO level (-5.24 eV vs. -5.47 eV) and a lower LUMO level (-1.59 eV vs. -1.03 eV) than HHTP, leading to a narrowed HOMO-LUMO gap (**Figure 2b**). The ESP map also shows a more anisotropic charge distribution across the molecular surface. This electronic restructuring confirms the formation of an extended, anisotropic π-conjugated system due to benzo-fusion, a conclusion supported by a redshift in the UV-vis absorption spectrum of BTH (290 nm) relative to HHTP (280 nm) (**Figure S8**). These results establish BTH as a site-differentiated ligand scaffold with graded acidity, electronic distribution, and coordination preference, which are essential for metal-selective topological assembly.

To examine the metal-selective assembly behavior of BTH, Cu(II) and Zn(II) were selected as representative divalent metal nodes. Initial screening showed that conventional solvothermal conditions commonly used for highly symmetric ligands such as HHTP failed to afford crystalline products, underscoring the synthetic challenges associated with the limited solubility and pronounced $pK_a$ gradient of BTH. The use of aqueous ammonia as a modulator proved essential, as it enhanced ligand dissolution while promoting the progressive deprotonation of hydroxyl groups according to their acidity hierarchy. Specifically, Cu-BTH-MOF was obtained as a dark-green solid in 80% yield by reacting BTH with copper(II) acetate monohydrate in a DMI/$H_2O$ mixed solvent system (3:1, v/v) containing aqueous ammonia at 65 °C for 24 h. Following an analogous procedure, Zn-BTH-MOF was prepared from BTH and zinc nitrate hexahydrate in a DMI/$H_2O$ solvent system (1:1, v/v; see **Schemes S2** and **S3** in the Supporting Information).

The structures of Cu-BTH-MOF and Zn-BTH-MOF were resolved by comparing the experimental PXRD patterns with the corresponding structural models, followed by Pawley refinement (**Figure S9**). Both frameworks were modeled using an AA eclipsed stacking arrangement. The experimental profiles showed excellent agreement with the simulated patterns, confirming the phase purity of the products and the reliability of the proposed structures. The refinement of Cu-BTH-MOF converged with $R_{wp}$ = 2.36% and $R_p$ = 1.88%, while Zn-BTH-MOF gave $R_{wp}$ = 4.21% and $R_p$ = 2.87% (**Figure 2c**), indicating the high quality of both structural models. Cu-BTH-MOF crystallizes in a trigonal system with unit-cell parameters of $a = b = c$ = 42.4254 Å, $\alpha = \beta$ = 90°, and $\gamma$ = 120°. The refined structure reveals a 2D framework featuring a dual-mode hexagonal pore topology, in which two chemically and geometrically distinct but topologically correlated pore environments coexist within an ordered lattice. In contrast, Zn-BTH-MOF crystallizes in an orthorhombic system with $a$ = 47.3112 Å, $b$ = 21.6645 Å, $c$ = 3.3700 Å, and $\alpha = \beta = \gamma$ = 90°, forming a uniform hexagonal channel topology (**Figure 2d**). This pronounced divergence in symmetry and pore architecture provides direct structural evidence that a single BTH ligand can encode multiple assembly pathways, which are selectively expressed through coordination with different metal ions.

The formation of metal-oxygen coordination environments in both BTH-based frameworks is confirmed by X-ray photoelectron spectroscopy (XPS). Survey spectra show the expected C, O, and metal signals, and high-resolution O 1s and metal core-level spectra are consistent with deprotonated BTH ligands coordinated to Cu and Zn nodes. Distinct core-level profiles further support the different local coordination environments of Cu-BTH-MOF and Zn-BTH-MOF (**Figures S10-S11**). Thermogravimetric analysis (TGA) also reveals distinct thermal responses. Cu-BTH-MOF shows a gradual weight-loss profile with a higher final residue of approximately 58-60%, whereas Zn-BTH-MOF exhibits a more pronounced weight-loss step and a lower residue of approximately 45-46% (**Figures S12-S13**). These differences suggest distinct guest-retention and decomposition behaviors, which may arise from their different pore environments. Nitrogen ($N_2$) adsorption at 77 K indicates permanent porosity. Cu-BTH-MOF and Zn-BTH-MOF exhibit BET surface areas of 328.8 and 35.0 $m^2\ g^{-1}$, respectively (**Figure 2g**). The corresponding pore size distributions further corroborate the structural models: Cu-BTH-MOF exhibits a broader and more complex distribution, in agreement with its dual-pore framework, whereas Zn-BTH-MOF shows a narrower distribution, indicative of a more uniform channel environment.

Preliminary morphological analysis by scanning electron microscopy (SEM) reveals distinct aggregation behaviors for the two BTH-based frameworks: Cu-BTH-MOF forms discoidal, plate-like particles, whereas Zn-BTH-MOF consists of flake-like subunits assembled into flower-like aggregates (**Figures S18-S19**). Cryo-transmission electron microscopy (cryo-TEM) provides direct real-space evidence of the internal lattice organization. Both Cu-BTH-MOF and Zn-BTH-MOF exhibit well-defined crystalline domains with sharp fast Fourier transform (FFT) patterns, confirming their long-range structural order. Notably, the local pore arrangements observed in the cryo-TEM images are fully consistent with the refined structural models. Cu-BTH-MOF displays heterogeneous pore environments characteristic of a dual-channel [444,46] framework, whereas Zn-BTH-MOF shows a more regular lattice consistent with a uniform [446] topology (**Figure 2f** and **2i**). These results provide microscopic support for the metal-dependent topological divergence of BTH-based MOFs.

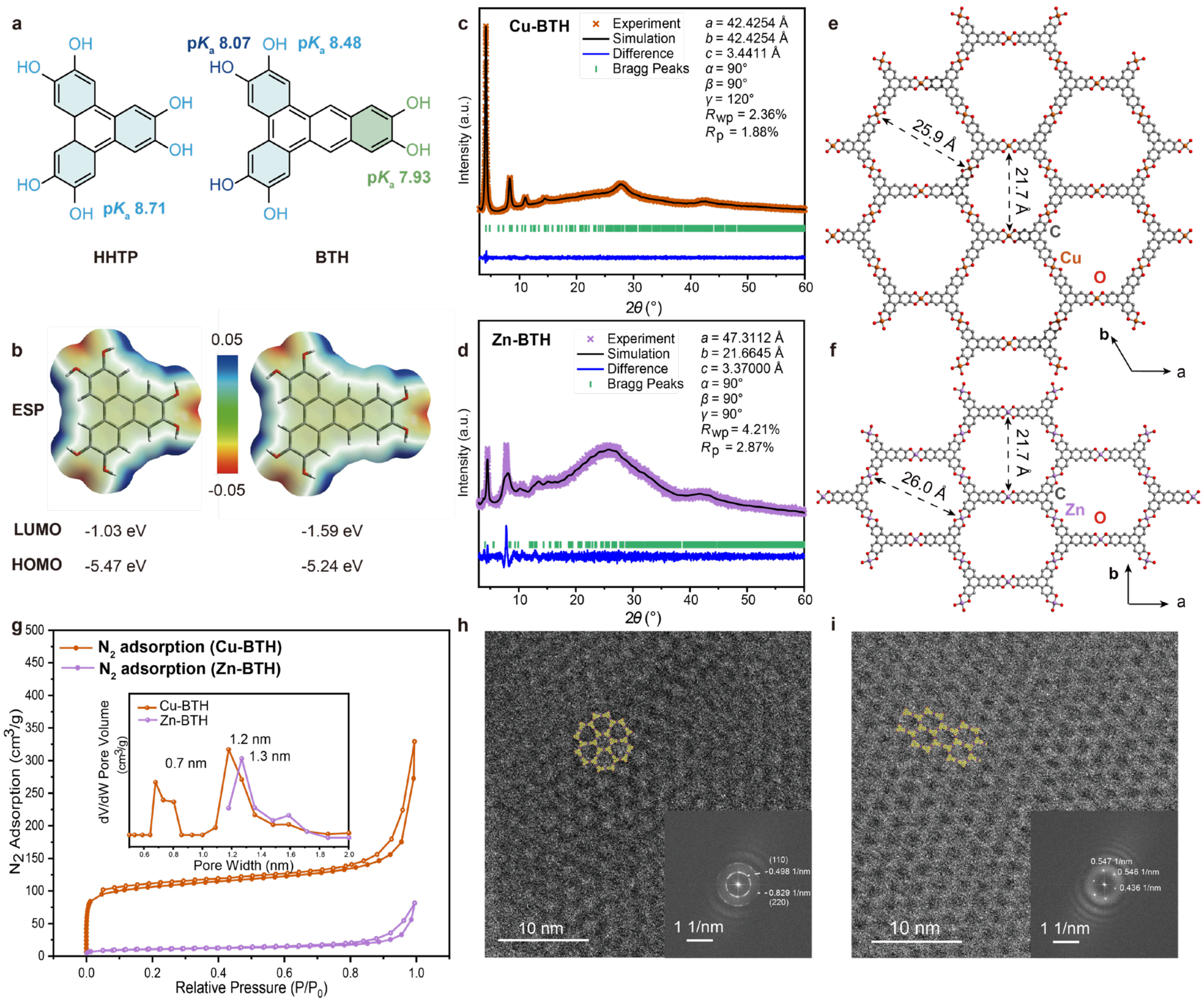

Figure 2. (a) Calculated p$K_a$ gradient and (b) electrostatic potential maps (ESP) and HOMO/LUMO energy levels of HHTP and BTH. (c, d) PXRD Pawley refinement profiles of Cu-BTH-MOF and Zn-BTH-MOF, confirming the proposed framework structures. (e,f) Structural models of Cu-BTH-MOF and Zn-BTH-MOF, with both structures refined under an AA eclipsed stacking model. (g) $N_2$ adsorption-desorption isotherms of Cu-BTH-MOF (shown in burnt orange) and Zn-BTH-MOF (shown in violet). The inset shows the corresponding pore size distribution curves. (h, i) Cryo-TEM images of Cu-BTH-MOF and Zn-BTH-MOF (Inset: fast Fourier transform (FFT) of images).

DFT calculations of binding energies for BTH–metal dimers provide a unified energetic framework for understanding the distinct topology formation mechanisms in Cu-BTH and Zn-BTH systems (**Figure 3**). The results demonstrate that, across both metals, the 3+3 dimer represents the most stable configuration (Cu-BTH: −5.7769 eV; Zn-BTH: −5.2617 eV), consistent with the p$K_a$ distribution and indicating preferential coordination on the benzo-fused arm of BTH. Importantly, the pronounced energetic differences between the *trans*-2+2 and *cis*-2+2 dimer configurations in the two metal systems are likely the key factor underlying the markedly different topological outcomes observed in the resulting MOFs. For Cu-BTH, the *cis*-2+2 motif (-5.3581 eV) is significantly more stable than the *trans*-2+2 motif (-4.8525 eV), establishing a pronounced energetic hierarchy among local coordination pathways (**Figure S22**). In particular, the relatively small energy difference between the *cis*-2+2 and 3+3 motifs facilitates their coparticipation in framework assembly. Combined with the intrinsically fast coordination kinetics of $Cu^{2+}$, this energetic landscape leads to the simultaneous incorporation of 3+3 and *cis*-2+2 motifs during network growth, ultimately favoring the formation of the non-equivalent dual-domain tiling structure [444,46].[31] In contrast, Zn-BTH exhibits nearly degenerate energies for the *trans*-2+2 (-4.8624 eV) and *cis*-2+2 (-4.8678 eV) motifs, indicating the absence of a strong energetic preference among secondary coordination pathways. Together with the slower coordination kinetics of $Zn^{2+}$, this results in the coexistence of multiple motifs during assembly, allowing the system to evolve toward a thermodynamically optimized global packing configuration. However, this energetic near-degeneracy alone is insufficient to fully rationalize the observed topological selectivity.

To resolve this, we calculated the average area per metal center in the [446] and [444,46] frameworks, corresponding to the effective area occupied by each node. Normalizing the binding energies by this area yields an energy density, which enables direct comparison of the thermodynamic stability of different topologies on a per-unit-area basis. The resulting stability landscape shows a clear topology dependence (**Figure S23**). For Cu-BTH, although the [444, 46] topology is intrinsically less favorable due to its larger nodal area and reduced packing density, the stronger intrinsic stabilization of the *cis*-2+2 motif compensates for this penalty, making the expanded topology energetically competitive. In contrast, for Zn-BTH, where the *trans*-2+2 and *cis*-2+2 motifs are nearly isoenergetic, the overall stability is primarily governed by packing efficiency, which favors the more compact [446] framework.

Overall, topology selection in BTH-based MOFs can be rationalized within a coupled energetic–kinetic–geometric (E–K–G) hierarchy involving three key factors: (i) a universal preference for 3+3 coordination as the initial assembly unit, (ii) metal-dependent variations in the binding energies of secondary coordination motifs, and (iii) topology-dependent energy density associated with extended assembly of frameworks. Together, these factors provide a qualitative framework of energetics that connects local coordination behavior with the metal-selective emergence of distinct 2D architectures. Within this framework, BTH functions as an encoded asymmetric ligand, in which site-differentiated coordination environments encode multiple possible assembly pathways. The identity of the metal center then determines which pathway is preferentially expressed during framework construction, ultimately giving rise to distinct 2D topological outcomes from a single molecular precursor.

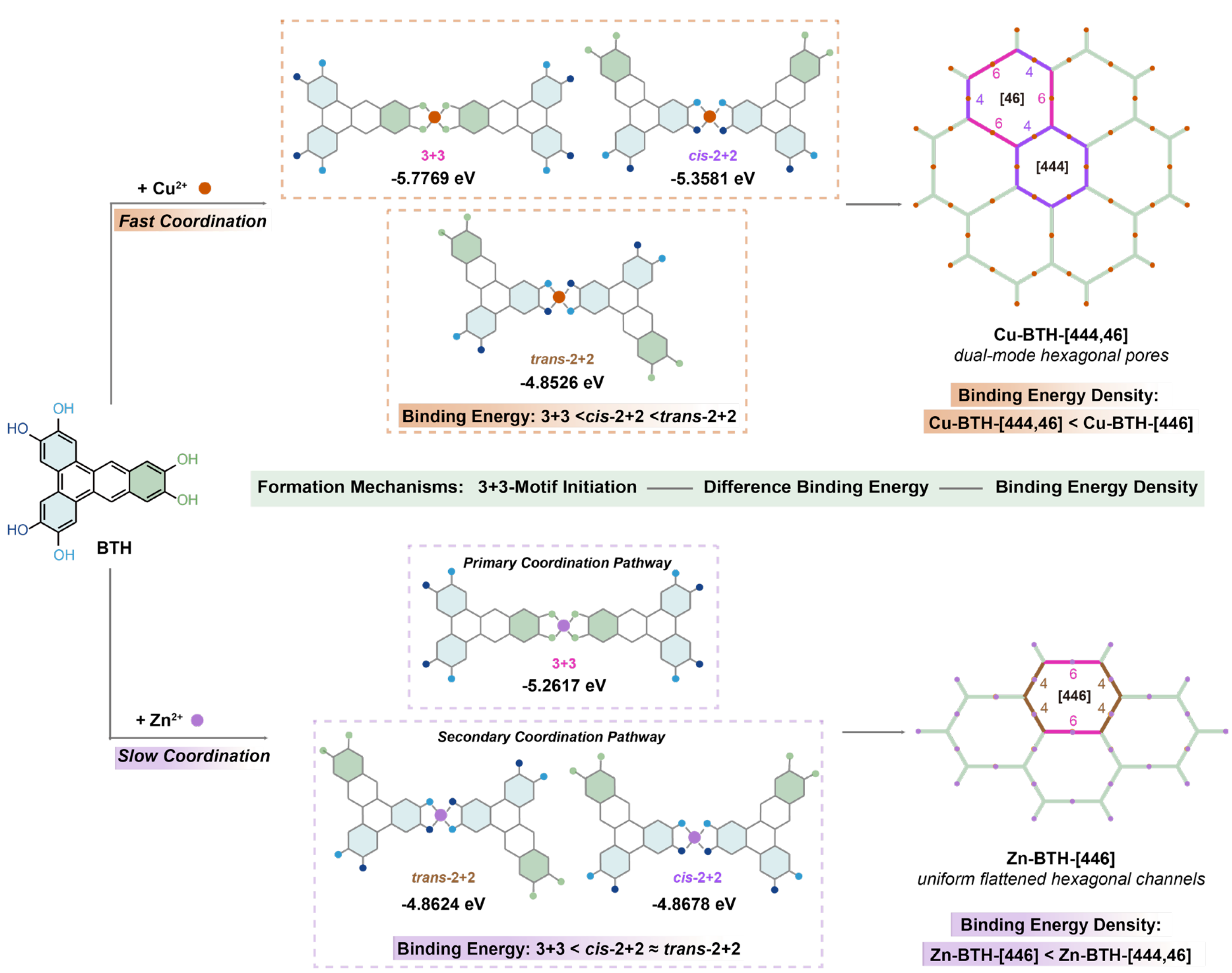

Figure 3. Metal-selective topological assembly of BTH-based 2D MOFs via Energetic–Kinetic–Geometric (*E*–*K*–*G*) Control. DFT calculations reveal a universal preference for the 3+3 motif in both Cu- and Zn-BTH systems, while secondary motifs exhibit metal-dependent energetic differences. In Cu-BTH, distinct motif energetics enable mixed 3+3/*cis*-2+2 assembly, leading to a dual-domain [444,46] topology. In contrast, Zn-BTH favors nearly degenerate motifs, yielding a uniform [446] framework. Energy normalization by metal-centered area defines a topology-dependent energy density that captures packing effects and rationalizes the distinct thermodynamic preference of the two systems.

Beyond the energetic and structural origins of topology selection, the metal-dependent topological divergence is also reflected in the electronic band structures of BTH-MOFs. DFT calculations reveal that the enlarged in-plane periodicity of the Cu-BTH-MOF topology compresses the Brillouin zone and flattens the in-plane band dispersion, thereby increasing the density of states (DOS) near the Fermi level (**Figure 4a**). In contrast, the Zn-BTH analogue exhibits a substantially wider band gap and diminished frontier-state density, owing to the closed-shell nature of $Zn^{2+}$ (**Figure 4b**). Although both materials show band dispersion along the Γ–A direction, the insulating characteristics of Zn-BTH for in-plane wavevectors suggest its lower electrical conductivity, as demonstrated below. To further emphasize the effect of ligand asymmetry, we computed the high-symmetry honeycomb Cu-HHTP framework as a reference (**Figure 4c**). Although both Cu-BTH and Cu-HHTP feature Cu–O coordination networks with finite DOS at the Fermi edge, the programmable BTH ligand, with its reduced symmetry and expanded real-space periodicity, gives rise to flatter in-plane bands and smaller effective band gaps compared with Cu-HHTP. Collectively, these results highlight the coupled influence of metal identity and ligand geometry on the electronic structures of BTH-based MOFs.

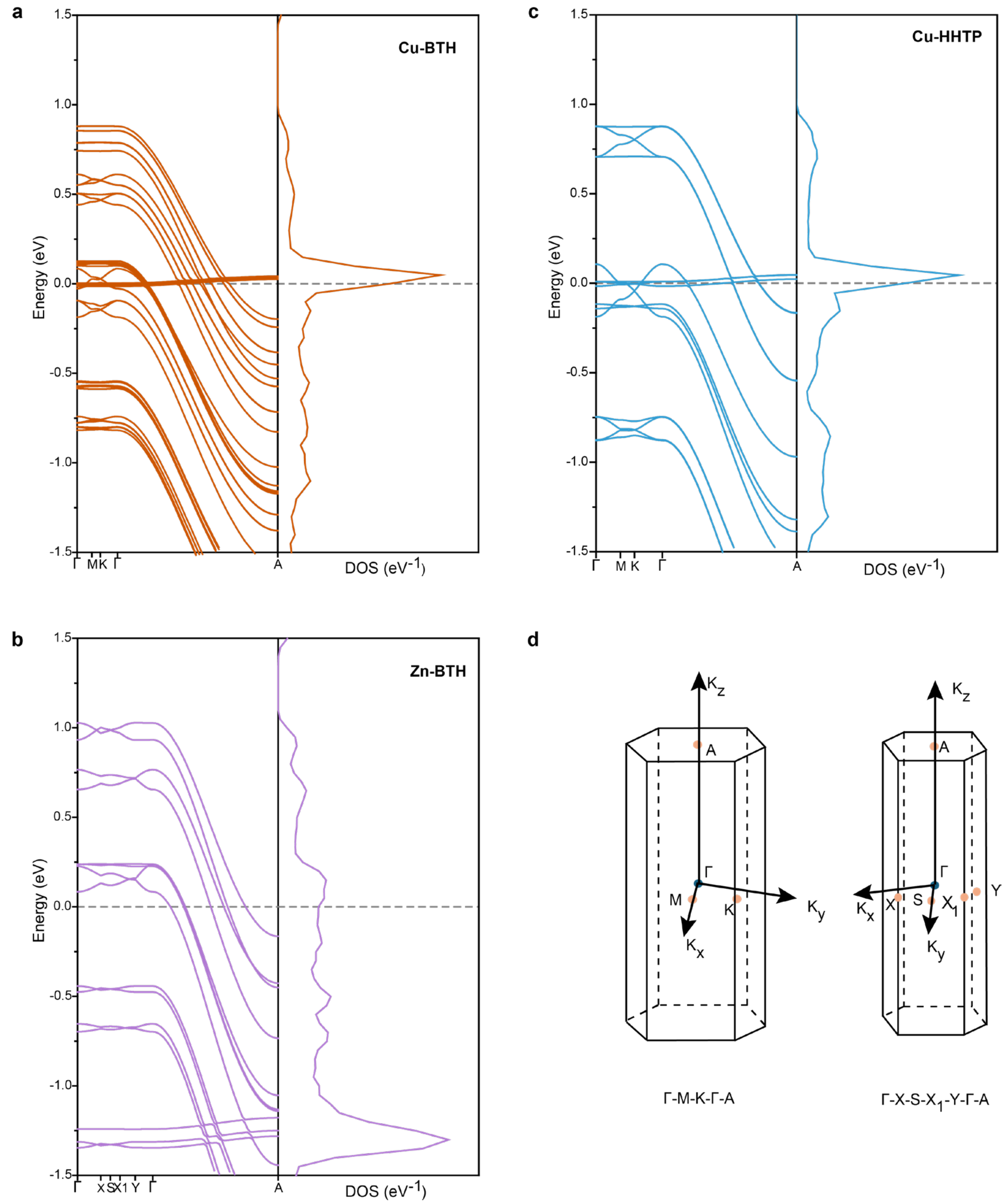

Figure 4. Calculated electronic structures of Cu-BTH-MOF, Zn-BTH-MOF, and Cu-HHTP. (a-c) DFT-calculated band structures and corresponding density of states (DOS) of Cu-BTH-MOF, Zn-BTH-MOF, and Cu-HHTP, respectively. The dashed horizontal lines denote the Fermi level. (d) Schematic comparison of the high-symmetry paths used for band-structure calculations in the corresponding Brillouin zones.

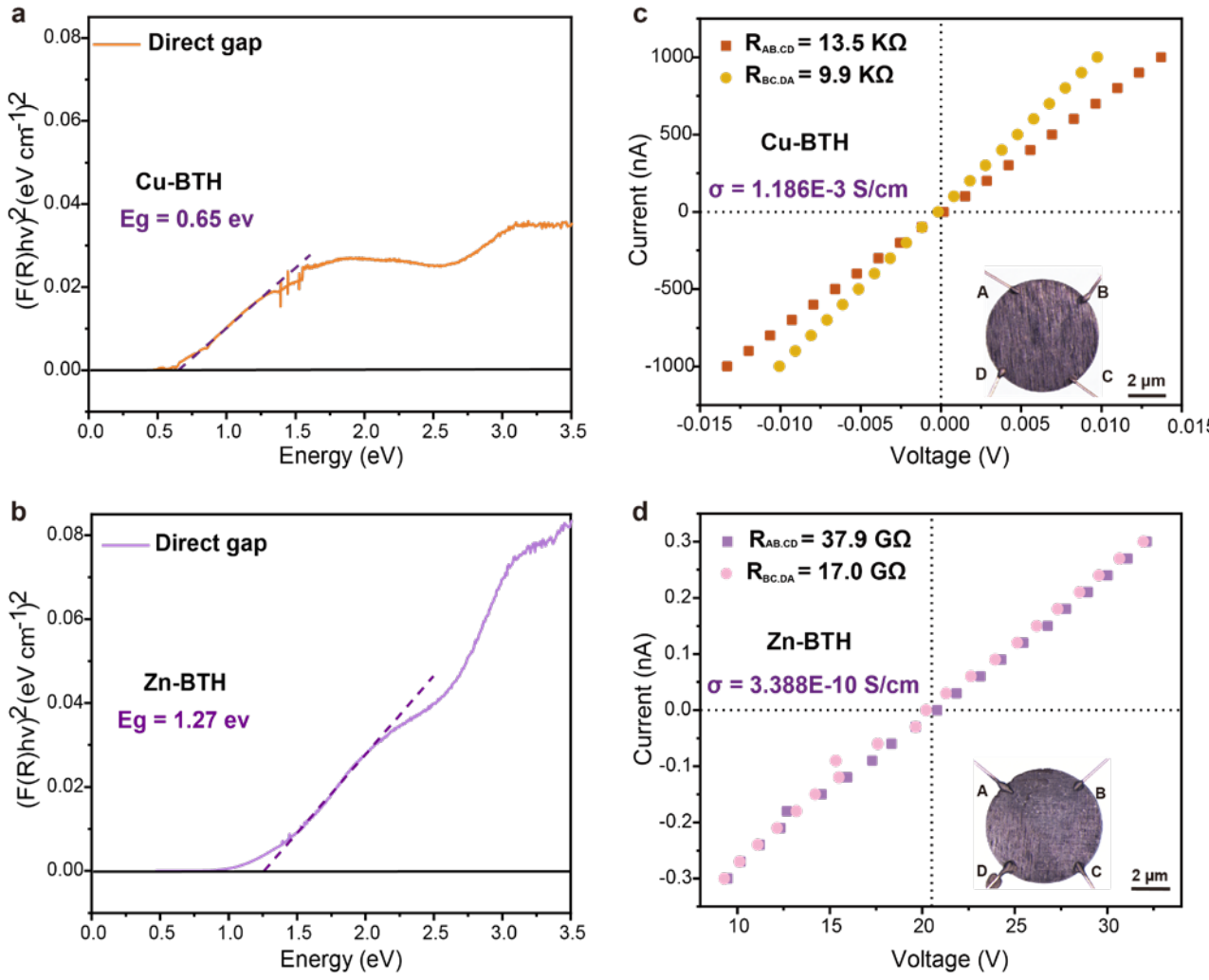

Figure 5. Electrical properties of Cu-BTH-MOF and Zn-BTH-MOF. (a, b) Tauc plots derived from diffuse-reflectance spectra of Cu-BTH-MOF and Zn-BTH-MOF. (c, d) Room-temperature electrical measurement of pressed powder pellets by the van der Pauw method.

We further measured the optical band gaps and macroscopic electrical conductivities of the two frameworks. Diffuse-reflectance measurements yield optical gaps of 0.65 eV for Cu-BTH-MOF and 1.27 eV for Zn-BTH-MOF, confirming that the two metal-selected topologies produce different band-edge transitions (**Figure 5a-b**). Powder van der Pauw measurements[32] further reveal a large difference in effective conductivity, with Cu-BTH-MOF reaching 1.186 × $10^{-3}$ S $cm^{-1}$ and Zn-BTH-MOF showing a much lower value of 3.38 × $10^{-10}$ S $cm^{-1}$ (**Figure 5c-d**). These results indicate that the metal-selective assembly of BTH does not merely alter the geometric arrangement of the framework; by changing the local coordination motifs, pore topology, and interlayer registry, it also reshapes frontier electronic states and orbital coupling pathways, ultimately leading to distinct optical gaps and macroscopic charge-transport properties.

### Conclusion

In summary, we have developed BTH as an encoded asymmetric ligand for metal-selective topology modulation in 2D MOFs. By introducing a single benzo-fused extension into the HHTP scaffold, BTH preserves the hexahydroxy coordination motif while creating site-differentiated coordination environments. This design allows one ligand to translate the identity of metal ions into distinct framework topologies, producing Cu-BTH-MOF with dual-mode hexagonal pores and Zn-BTH-MOF with uniform flattened hexagonal channels. Structural, porosity, microscopy, and spectroscopic analyses confirm the one-ligand, divergent-topology behavior. DFT calculations, diffuse-reflectance spectroscopy, and powder conductivity measurements show that the metal-directed topologies endow the resulting frameworks with unique and markedly different electronic structures, optical gaps, and macroscopic charge-transport properties. The conductivity of Cu-BTH-MOF (1.186 × $10^{-3}$ S $cm^{-1}$) exceeds that of Zn-BTH-MOF (3.38 × $10^{-10}$ S $cm^{-1}$) by more than six orders of magnitude. This work establishes programmable asymmetric ligand design as a general strategy for encoding multiple MOF topologies from a single molecular scaffold while linking metal-selective framework assembly with tunable electronic functionalities.

## ASSOCIATED CONTENT

### Supporting Information

The following files are available free of charge.
Complete experimental procedures and other supporting data. (PDF)

## AUTHOR INFORMATION

### Corresponding Author

*Jin-Hu Dou
doujinhu@pku.edu.cn (J. D.)
*Qingqing Ji
jiqq@shanghaitech.edu.cn (Q. J.)

### Author Contributions

†Huimin Qi, Jinkun Guo, and Xinyan Wu contributed equally to this work.

### Funding Sources

This work was financially supported by the National Key Research and Development Program of China (Grant Nos. 2023YFE0206400 and 2024YFA1410500), National Natural Science Foundation of China (Grant Nos. 22171185, 22575005), and the Science and Technology Commission of Shanghai Municipality (Grant No. 25JC3200500).

### Notes

The authors declare no competing interests.

### ACKNOWLEDGMENT

This work was financially supported by the National Key R&D Program of China (Grant No. 2023YFE0206400), National Natural Science Foundation of China (Grant No. 22171185, 22575005). Cryo-EM was performed at Electron Microscopy Laboratory of Peking University with the assistance of Xuemei Li. We appreciate Beijing Nuclear Magnetic Resonance Center for NMR characterization with the assistance of Hongwei Li and Xiaogang Niu.

## REFERENCES

(1) Hou, L.; Cui, X.; Guan, B.; Wang, S.; Li, R.; Liu, Y.; Zhu, D.; Zheng, J. Synthesis of a monolayer fullerene network. *Nature* **2022**, *606*, 507–511.
(2) Cao, Y.; Fatemi, V.; Fang, S.; Watanabe, K.; Taniguchi, T.; Kaxiras, E.; Jarillo-Herrero, P.; Unconventional superconductivity in magic-angle graphene superlattices. *Nature* **2018**, *1556*, 43–50.
(3) Guo, Q.; Zhang, Q.; Zhang, T.; Zhou, Jun.; Xiao, S.; Wang, S.; Feng, Y. P.; Qiu, C.-W.; Colossal in-plane optical anisotropy in a two-dimensional van der Waals crystal. *Nat. Photon.* **2024**, *18*, 1170–1175.
(4) Xie, L. S.; Skorupskii, G.; Dincă, M. Electrically Conductive Metal−Organic Frameworks. *Chem. Rev.* **2020**, *120*, 8536−8580.
(5) Zhou, H.-C.; Long, J. R.; Yaghi, O. M. Introduction to Metal−Organic Frameworks. *Chem. Rev.* **2012**, *112*, 673−674.
(6) Wang, M.; Dong, R.; Feng, X. Two-dimensional conjugated metal−organic frameworks (2D c-MOFs): chemistry and function for MOF tronics. *Chem. Soc. Rev.* **2021**, *50*, 2764−2793.
(7) Chakraborty, G.; Park, I. H.; Medishetty, R.; Vittal, J. J. Two Dimensional Metal-Organic Framework Materials: Synthesis, Structures, Properties and Applications. *Chem. Rev.* **2021**, *121*, 3751−3891.
(8) Li, C.; Wang, Y.; Jiang, Y.; Liu, Y.; Wu, Y.; Han, Y.-J.; Chen, Z.-

Q.; Liu, L.; Liu, X.; Guan, D.; Li, Y.; Zheng, H.; Liu, C.; Liu, P.-N.; Jia J.; Li, D.-Y.; Wang, S. Programmable Higher-Order Topological Phases in Open-Shell Metal−Organic Frameworks. *J. Am. Chem. Soc.* **2025**, *147*, 39662−39670.
(9) Liu, J. J.; Xing, G. L.; Chen, L. 2D Conjugated Metal−Organic Frameworks: Defined Synthesis and Tailor-Made Functions. *Acc. Chem. Res.* **2024**, *57*, 1032−1045.
(10) Dou, J.-H.; Arguilla, M. Q.; Luo,Y.; Li, J.; Zhang,W.; Sun, L.; Mancuso, J. L.; Yang, L.; Chen, T.; Parent, L. R.; Skorupskii, G.; Libretto, N. J.; Sun, C.; Yang, M. C.; Dip, P. V.; Brignole, E. J.; Miller, J. T.; Kong, J.; Hendon, C. H.; Sun J.; Dinca, M. Atomically precise single-crystal structures of electrically conducting 2D metal–organic frameworks. *Nat. Mater.* **2021**, *20*, 222−228.
(11) Liu, Y.; Yao H.; Zhang H.; Ma C.; Guo J.; Fan Y.; Chen H.; Wu X.; Yang L.; Huang X.; Chen T.; Ji Q.; Yao Z.-F.; Li J.; Dou J.-H. Asymmetrical Substitution Manipulates Stacking Modes in 2D Conductive MOF Crystals. *J. Am. Chem. Soc.* **2025**, *147*, 48127−48135.
(12) Fan,Y.; Jiang, B.; Zhang, Z.; Zhang, H.; Yang, L.; Chen, T.; He, L.; Liu,Y.; Guo, J.; Zhao, T.; Du, R.; Tang, C.; Li, J.; Zheng, M.; Dou, J.-H. Packing Order Control in Conductive Metal−Organic Frameworks by Tuning Ligand Oxidation State. *Angew. Chem. Int. Ed.* **2026**, *65*, 2−8.
(13) Qi, M. L.; Zhou, Y.; Lv, Y. K.; Chen, W. B.; Su, X.; Zhang, T.; Xing, G. L.; Xu, G.; Terasaki, O.; Chen, L. Direct Construction of 2D Conductive Metal−Organic Frameworks from a Nonplanar Ligand: In Situ Scholl Reaction and Topological Modulation. *J. Am. Chem. Soc.* **2023**, *145*, 2739−2744.
(14) Qing, H.; Chandra, P.; Chan, J. Y. M.; Staples, R. J.; Li, T.-D.; Li, W.; Mirica, K. A. Topological Control of Dual Protonic–Electronic Conduction in Metal–Organic Frameworks. *J. Am. Chem. Soc.* **2026**, *148*, 28742−28753.
(15) Wang, Z.; Ye, X.; Zhang, Y.; Yang, Q.; Gao, K.; Li, Z.; Zhuo, H.; Wang, Z.; Liu, J.; Yuan, H.; Fan, K.; Meng, Z.; Shang, X.; A naphthacene-based two-dimensional conductive metal–organic framework for highly efficient chemiresistive sensing of ammonia. *J. Mater. Chem. A* **2025**, *13*, 27115–27124.
(16) Yang, M.; Zhang, Y.; Zhu, R.; Tan, J.; Liu, J.; Zhang, W.; Zhou, M.; Meng, Z.; Two-Dimensional Conjugated Metal–Organic Frameworks with a Ring-in-Ring Topology and High Electrical Conductance. *Angew. Chem. Int. Ed.* **2024**, *63*, e202405333.
(17) Feng, D. W.; Wang, K. C.; Su, J.; Liu, T. F.; Park, J.; Wei, Z. W.; Bosch, M.; Yakovenko, A.; Zou, X. D.; Zhou, H.-C. A Highly Stable Zeotype Mesoporous Zirconium Metal−Organic Framework with Ultralarge Pores. *Angew. Chem., Int. Ed.* **2015**, *54*, 149−154.
(18) Feng, L.; Wang, Y. T., Zhang, K.; Wang, K.-Y.; Fan, W. D.; Wang, X. K.; Powell, J. A.; Guo, B. B.; Dai, F. N.; Zhang, L. L.; Wang, R. M.; Sun, D. F.; Zhou, H.-C. Molecular Pivot-Hinge Installation to Evolve Topology in Rare-Earth Metal−Organic Frameworks. *Angew. Chem. Int. Ed.* **2019**, *58*, 16682−16690.
(19) Lv, X.-L.; Feng, L.; Xie, L. H.; He, T.; Wu, W.; Wang, K.-Y.; Si, G. R.; Wang, B.; Li, J.-R.; Zhou, H.-C. Linker Desymmetrization: Access to a Series of Rare-Earth Tetracarboxylate Frameworks with Eight-Connected Hexanuclear Nodes. *J. Am. Chem. Soc.* **2021**, *143*, 2784−2791.
(20) Han, W. T.; Ma, X.; Wang, J. X.; Leng, F. C.; Xie, C. F.; Jiang H.-L. Endowing Porphyrinic Metal−Organic Frameworks with High Stability by a Linker Desymmetrization Strategy. *J. Am. Chem. Soc.* **2023**, *145*, 9665−9671.
(21) Alezi, D.; Peedikakkal, A. M. P.; Weseliński, L. J.; Guillerm, V.; Belmabkhout, Y.; Cairns, A. J.; Chen, Z.; Wojtas, L.; Eddaoudi, M. Quest for Highly Connected Metal−Organic Framework Platforms: Rare-Earth Polynuclear Clusters Versatility Meets Net Topology Needs. *J. Am. Chem. Soc.* **2015**, *137*, 5421−5430.
(22) Guillerm, V.; Grancha, T.; Imaz, I.; Juanhuix, J.; Maspoch, D. Zigzag Ligands for Transversal Design in Reticular Chemistry: Unveiling New Structural Opportunities for Metal−Organic Frameworks. *J. Am. Chem. Soc.* **2018**, *140*, 10153−10157.
(23) Guillerm, V.; Maspoch, D. Geometry Mismatch and Reticular Chemistry: Strategies to Assemble Metal−Organic Frameworks with Non-default Topologies. *J. Am. Chem. Soc.* **2019**, *141*, 16517−16538.
(24) Chen, P.; Liu, M.; Li, R.; Su, X.; Xing, G.; Ren, X.-R.; Wang, D.; Zhang, J.; Chen, L. Multidentate Macrocyclic Salphen-Based 2D Conjugated Metal−Organic Framework. *Angew. Chem. Int. Ed.* **2025**, *64*, e202511048.
(25) Dinakar, B.; Oppenheim, J. J.; Vandone, M.; Torres, J. F.; Iliescu, A.; Yang, Z.; Roma´n-Leshkov, Y.; Dinca, M. Asymmetric linker generates intrinsically disordered metal–organic framework with local MOF-74 structure. *Chem. Commun.* **2025**, *61*, 12590–12593.
(26) Avinash, K.; Clifford II, W. C.; Dye, A. R.; Adeniyi, A. E.; Everitt, K. L.; Streblow, G. J.; Poore, A. T.; Jager, A. K.; Tian, S.; Andrews, J. L. Constructing a Two-Dimensional Electrically Conductive Metal−Organic Framework from a Ligand with Low In-Plane Rotational Symmetry. *Chem. Mater.* **2026**, *38*, 6059−6068.
(27) Cheng, Y.-Z.; Kong, H.-Y.; Hao, P.-Y.; Huang, K.; Ding, X. S.; Shi, X. H.; He, Y. J.; Han, B.-H. Reticular Synthesis of Covalent Organic Frameworks with kgd-v Topology and Trirhombic Pores. *J. Am. Chem. Soc.* **2025**, *147*, 4844−4852.
(28) Zasada, L. B.; Guio, L.; Kamin, A. A.; Dhakal, D.; Monahan, M.; Seidler, G. T.; Luscombe, C. K.; Xiao, D. J. Conjugated Metal−Organic Macrocycles: Synthesis, Characterization, and Electrical Conductivity. *J. Am. Chem. Soc.* **2022**, *144*, 4515−4521.
(29) Wei, J.; Han, B.; Guo, Q.; Shi, X.; Wang, W.; Wei, N.; 1,5,9-Triazacoronenes: A Family of Polycyclic Heteroarenes Synthesized by a Threefold Pictet–Spengler Reaction. *Angew. Chem. Int. Ed.* **2010**, *49*, 8209–8213.
(30) Meng, Z.; Mirica, K. A.; Two-dimensional d-π conjugated metal−organic framework based on hexahydroxytrinaphthylene. *Nano Res.* **2021**, *14*, 369−375.
(31) Chen, H.; Zhao, T.; Li, W.; He, L.; Guo, J.; Yao, Z.-F.; Ma, C.; Fan, Y.; Jiang, B.; Zhang, L.; Zhang, H.; Liu, S.; Zhang, Z.; Sun, L.; Yang, L.; Brozek, C. K.; Zheng, Y.-Q.; Li, J.; Chen, T.; Sun, J.; Dou, J.-H. Aligning Chemical Kinetics with Crystallization Enables Millimeter Scale Single Crystals of Conductive MOFs. *J. Am. Chem. Soc.* **2026**, *148*, 27710−27721.
(32) Sun, L.; Park, S. S.; Sheberla, D.; Dinca, M. Measuring and Reporting Electrical Conductivity in Metal−Organic Frameworks: $Cd_2$(TTFTB) as a Case Study. *J. Am. Chem. Soc.* **2016**, *138*, 14772−14782.

TOC

**An Encoded Asymmetric Ligand Enables Metal-Selective Topological Assembly in 2D MOFs**

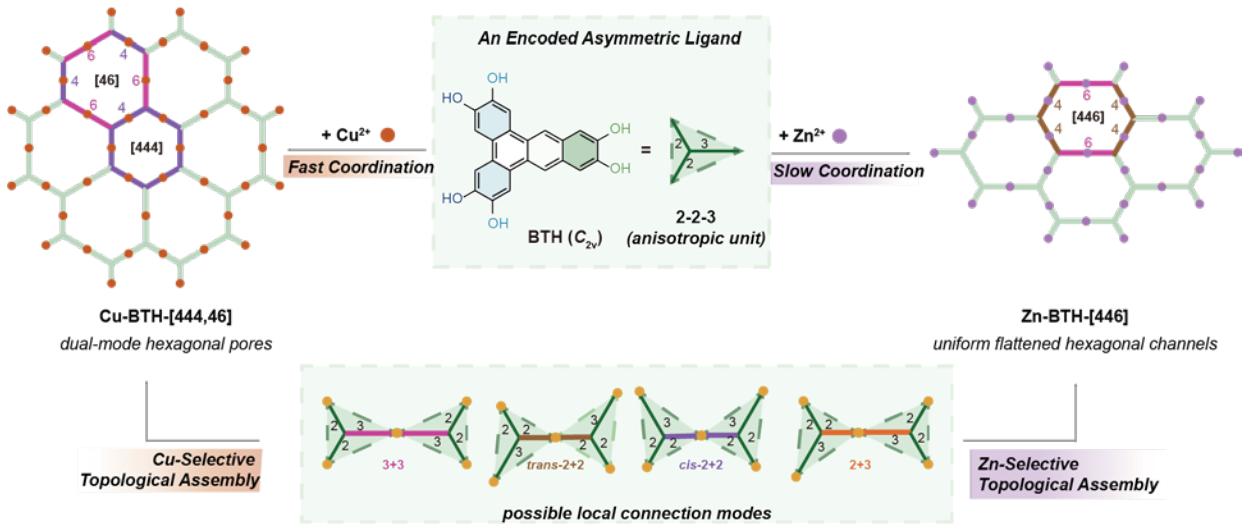